\documentclass{article}
\usepackage{geometry}
\usepackage{graphicx}
\usepackage{subfigure}
\usepackage{amssymb}
\usepackage{amsmath}
\usepackage{mathrsfs}
\usepackage{epstopdf}
\usepackage{cite}
\usepackage{bbm}
\usepackage{multirow}
\usepackage{enumerate}
\usepackage{xcolor}
\usepackage[hyperindex=true,
          pdfstartview=FitH,
          bookmarksnumbered=true,
          bookmarksopen=true,
          citecolor=blue,
          linkcolor=blue,
          colorlinks=true,
          pdfborder=001,
          unicode]{hyperref}

\newcommand{\ba}{\begin{align}}
\newcommand{\ea}{\end{align}}
\newcommand{\be}{\begin{equation}}
\newcommand{\en}{\end{equation}}
\newcommand{\bea}{\begin{eqnarray}}
\newcommand{\ena}{\end{eqnarray}}

\usepackage{graphicx} 

\title{Gravitational waves of extreme-mass-ratio inspirals in a rotating black hole with Dehnen dark matter halo}
\author{Kun Meng$^1$\footnote{kunmeng@wfu.edu.cn}, Shao-Jun Zhang$^{2,3}$\footnote{sjzhang@zjut.edu.cn}, Nan Yang$^4$\footnote{201610990@xttc.edu.cn}\\
$^1$School of Physics and Electronic Information,
	Weifang University, Weifang 261061, China \\
$^2$Institute for Theoretical Physics \& Cosmology,\\
Zhejiang University of Technology, Hangzhou 310032, China\\
 $^3$United Center for Gravitational Wave Physics,\\
Zhejiang University of Technology, Hangzhou 310032, China
    \\
 $^4$Department of Electronical Information Science and Technology, \\
 Xingtai University, Xingtai 054001, China
}

\begin{document}

\maketitle

\begin{abstract}
Extreme Mass Ratio Inspirals (EMRIs) are among the key target sources for the space-based gravitational wave (GW) detectors. The waveforms of the EMRIs are highly sensitive to the types of the central supermassive black hole (SBH) and can serve as a novel sensitive tool to probe the background spacetime. In this work,  we compute GWs radiated from EMRIs in the backgrounds of Kerr black hole and rotating black hole with Dehnen-type dark matter halo (DMBH). Following the Teukolsky prescription, we obtain the perturbed equations for curvature tensor from the Newman–Penrose (NP) equations, and for the DMBH we obtain the radial and angular equations through  separation of variables. To solve the equations with numerical method we apply the Sasaki-Nakamura (SN) transformation to convert the Teukolsky-type equation into the SN equation. We study the radiation reaction of GWs by computing the energy flux and angular momentum flux at infinity and at the horizon. The orbital evolution is then derived from the total fluxes.  We extract the two polarizations of GWs by solving the equation numerically. By comparing the waveforms of Kerr and DMBH, it is found that the dark matter halo induces noticeable changes in both the amplitude and phase of GWs. We compute the strain of GW detector with the response function and evaluate the mismatch between the waveforms of Kerr and DMBH. The results show that the mismatch increases with the mass parameter of DM halo and the spin of the SBH.
\end{abstract}

\section{Introduction}
\label{section1}
The discovery of GWs marks a major breakthrough in the field of gravitation\cite{LIGOScientific:2016aoc,LIGOScientific:2016sjg}. It provides strong evidence supporting the validity of Einstein's theory of gravity and opens a new era of detecting the universe. GWs can serve as a novel tool to probe the strong-field properties of black holes, the ``no-hair theorem'', fundamental fields, and the underlying theory of gravity.
Ground-based GW detectors have detected over two hundred GW events, including binary black hole mergers, binary neutron star mergers, and black hole-neutron star mergers. However, ground-based GW detectors can only detect GWs within specific frequency band, which ranges from $10$ to $10^4$ Hz. EMRIs are important sources of GWs, producing GWs with frequencies in the range from $10^{-4}$ to $10^{-1}$ Hz, this frequency band lies outside the range accessible by ground-based GW detectors.

EMRIs are one of the primary targets of the next-generation space-based GW detectors (such as Taiji, Tianqin, and LISA), it marks the beginning of the stage of precision measurement in GW detections. The detection of GWs emanating from EMRIs will enhance our understanding of gravitational theory, black holes, cosmic evolution, etc.  An EMRI system consists of a  SBH that is millions of times of the mass of the Sun and a stellar-mass compact object orbiting the central SBH\cite{Sigurdsson:1997vc,Sigurdsson:1996uz}. The small secondary in an EMRI system acts as an highly sensitive probe, its orbital motion depends exquisitely on the spacetime geometry around the SBH, so detections of GWs from EMRIs allow to map the spacetime near the black hole with unprecedented precision. The small secondary will execute tens of thousands of orbital cycles before eventually plunging into the event horizon of the SBH, so the GW signals from EMRI last for months to years, this prolonged ``song'' allows to detect the mass, spin or other characteristic parameters of the central SBH with remarkable accuracy.
EMRI events occur in the extreme regions of strong gravity. This provides excellent opportunities to probe the strong-field properties and multipole structure of black holes, as well as to test GR with high precision in the strong-field regime\cite{Hughes:2001jr,Glampedakis:2002ya,Gupta:2022fbe,Polcar:2022bwv,Munna:2022xts,Kerachian:2023oiw,Takahashi:2023flk,Burko:2013cca,Glampedakis:2005hs,Babak:2017tow,Munna:2019fjz,Destounis:2022obl,Barack:2003fp,Dai:2023cft,Pan:2023wau,Meng:2024kug,Santos:2025ass,Cocco:2025udb,Cocco:2025adu,Yin:2024nyz}.
EMRIs events are closely related to the dynamics of stellar-mass compact objects in galactic nuclei, detecting the GWs of EMRIs provides critical insights into the stellar distribution around galactic centers and the history of galaxy evolution\cite{Zhang:2025jmm,Liu:2026dug}.
EMRIs also serve as an ideal laboratory for testing fundamental physics, GWs emanating from EMRIs make it possible to detect new fundamental fields, probe the astronomical environment, and constrain existing theories\cite{Gair:2012nm,Will:1977wq,Will:1989sk,Barausse:2015wia,Xu:2021kfh,Tan:2024utr,
Gair:2011ym,Zhang:2022rfr,Liang:2022gdk,Rahman:2022fay,Qiao:2024gfb,Meng:2025glf,Xia:2026aty,Gong:2025mne,Xia:2025yzg,Lu:2025xlp,Li:2025sfe,Bonga:2019ycj,Li:2025ffh,Grilli:2024fds}, etc.

If the mass of a galaxy is supposed to be dominated by visible matter in the central region, it's expected that the velocities of stellar-mass objects orbiting the galactic center decrease with distance. However, observations reveal that beyond the optical size of galaxies, the velocities of stellar-mass objects do not fall down with distance but tend to be constant. This implies the existence of invisible DM within galaxies. DM constitutes one of the biggest mysteries nowadays, detecting and understanding DM have become one of the most important subjects in physics \cite{Drukier:1986tm,Dalal:2001fq,Abazajian:2001vt,Schneider:1996ug,Pierce:2018xmy}. Before GWs were discovered, the detection of the universe relied primarily on electromagnetic means, although many astronomical processes do not emit electromagnetic waves. GWs make it possible to observe the ``dark'' events that were inaccessible previously through traditional methods. On the distribution of DM halos around black holes, there is a debate between the cuspy and cored profiles. Based on the standard cold dark matter model, numerical simulations give rise to cuspy profile, whereas some astronomical observations reveal that DM in galactic centers are of cored-type. Some mechanisms such as baryonic feedback and self-interation have been introduced to explain the cored profile of DM\cite{DelPopolo:2021bom,Tulin:2017ara}. We will focus on the Dehnen profile of DM halo since it is flexible enough to admit both cored and cuspy configurations\cite{Dehnen:1993uh,Pantig:2022whj}. Due to the high-precision detection ability of EMRI GWs, they may be used as a novel sensitive tool to detect DM. If a black hole is surrounded by DM halo, the geometry of spacetime will deviate from the vacuum case, the orbits of the stellar-mass objects moving in this spacetime background will deviate from the vacuum case as well. The deviations would be encoded in the GWs. In this paper, we will explore the deviation from Kerr geometry induced by DM halo, and the subsequent deformation of gravitational waveforms of EMRIs. We will calculate the EMRI GWs in both Kerr and DMBH backgrounds, and make a comparison between the two types of waveforms to detect or constrain the characteristic parameters of DM halo.

Analysis of the data released by GW detectors replies heavily on the matched filtering technique, which makes correlations between data and theoretical waveforms of EMRIs\cite{Drasco:2005kz,Hughes:1999bq,Cutler:1994pb,Tanaka:1993pu,Shibata93,Tagoshi:1995sh,Poisson:1993vp,Cutler:1993vq, Apostolatos:1993nu,Poisson:1993zr,Poisson:1994yf,Tagoshi:1994sm,Shibata:1994jx,Cui:2026qsk,Jiang:2025mna,Zou:2025fsg,Shen:2025svs,Chen:2023ese}. Detections of theoretical parameters with GWs require high-precision waveforms. Among the methods to generate gravitational waveforms, the Teukolsky method is believed to be a valid one to generate high-precision waveforms\cite{Teukolsky:1973ha,Press:1973zz,Teukolsky:1974yv}.  Since recent observations support black holes possess spin\cite{LIGOScientific:2025rid},  this prompts us to study EMRI GWs of Kerr and rotating DMBH with Teukolsky method.Teukolsky method is based on Newman-Penrose (NP) formulation to establish the perturbation equations of background spacetime\cite{Newman:1961qr}.
For static black holes, one can establish perturbation equations for the metric. Odd-parity metric perturbations correspond to the Regge–Wheeler equation, and even-parity metric perturbations correspond to the Zerilli equation. For rotating black holes, however, establishing perturbation equations for the metric is very difficult. Typically, one first establishes perturbation equations for the curvature tensor, then solves the equations to extract the metric perturbations.
For DMBH, we follow this procedure: first establish the perturbation equations for the curvature, then solve them and extract the metric perturbations.
For the DMBH we set up the Teukolsky-type perturbation equations for $\psi_0$ and $\psi_4$ with the NP equations, and then separate variables for the perturbation equations to get the radial equation and angular equation respectively.  We solve the Teukolsky-like equations numerically to read off the two polarization states of GWs. Finally, we make comparisons between the gravitational waveforms of Kerr and DMBH to examine the imprints of deviation from Kerr in GWs. By adjusting the DM and spin parameters of the SBHs and calculating the mismatch of the two types of waveforms, we examine the correlation of the mismatch with these parameters.

The paper is organised as, in Section \ref{section2}, we establish the Teukolsky-type equations for curvature tensor perturbations and separate variables to obtain the radial and angular equations. Then, via the SN transformation, we transform the Teukolsky equations into the SN equations. In Section \ref{section3}, we study the orbital motion of the GW source, compute the radiation-reaction energy flux and angular momentum flux, and solve the equations numerically. In Section \ref{section4}, we calculate the detector’s strain through the response function, and then compare the waveforms of the Kerr black hole and the DMBH by computing the mismatch. We summarize our results in the last section.
In this work, the geometrized units are used with $G=c=1$.

\section{Master Equation\label{section2}}
\subsection{Teukolsky Equation}
The metric for the rotating black hole with Dehnen DM halo, which is obtained from the static counterpart through Newman-Janis prescription, is given by\cite{Pantig:2022whj}
\begin{eqnarray}\label{DMBH}
ds^2=-\left(1-\frac{2\mathcal{M}r}{\Sigma}\right)dt^2+\frac{\Sigma}{\Delta}dr^2-2a\sin^2\theta \frac{2\mathcal{M}r}{\Sigma} d\phi dt +\Sigma d\theta^2+\sin^2\theta\left[r^2+a^2+a^2\sin^2\theta \frac{2\mathcal{M}r}{\Sigma} \right]d\phi^2\ ,
\end{eqnarray}
where
\begin{equation}
\begin{aligned}\label{Delta}
\Delta&=r^2-2\mathcal{M}r+a^2,\\
\mathcal{M}&=\frac{r}{2}\left[1-\exp\left(-\frac{2\tilde{k} r^{\tilde{\sigma}-2} (r_c+r)^{\tilde{\sigma}-2}}{r_c(\tilde{\sigma}-2)}\right)\right]+M,
\end{aligned}
\end{equation}
here the parameters $\tilde{k}$ and $r_c$ denotes the DM halo's total mass and the scale radius of the DM halo, respectively. The DM parameter $\tilde{\sigma}$ is restricted to the range $[0,2)$ to ensure the line element (\ref{DMBH}) describes a stationary black hole. $\tilde{\sigma}=0$ describes a cored profile of DM halo, while $\tilde{\sigma}=1$ describes a cuspy one.

Since it is very difficult to directly establish perturbation equations for the metric for rotating black holes, a feasible approach is to first establish the perturbation equations for the curvature tensor, and then take the limit  $r\rightarrow\infty$ to obtain the perturbations for the metric. Following the prescription of Kerr, one establishes the perturbation equations for curvature tensor with the NP formulation for the DMBH (\ref{DMBH}).
The involved NP equations can be written in a simplified form\cite{Li:2022pcy,LaHaye:2025ley}
\begin{align}
F_1\psi_0-J_1\psi_1-3\kappa \psi_2&=\hat{T}_0,\label{NP1}\\
F_2\psi_0-J_2\psi_1-3\sigma \psi_2&=\tilde{T}_0,\label{NP2}\\
E_2\sigma-E_1\kappa-\psi_0&=0,\label{NP3}
\end{align}
by introducing the operators
\begin{equation}
\begin{aligned}
F_1&=\delta^*-4\alpha+\pi,\quad\quad\quad\quad\quad F_2=\Delta_0+\mu-4\gamma,\\
J_1&=D-2\varepsilon-4\rho,\quad\quad\quad\quad\quad J_2=\delta-2\beta-4\tau,\\
E_1&=\delta-\tau+\pi^*-\alpha^*-3\beta,\quad E_2=D-\rho-\rho^*-3\varepsilon+\varepsilon^*,
\end{aligned}
\end{equation}
where the derivative operators, spin coefficients, tensor components and source terms used here are defined identically with that given in Ref.\cite{Teukolsky:1973ha}. Notice that Eqs.(\ref{NP1}) and (\ref{NP2}) are coupled equations for curvature tensors. To decouple the equations and single out the equation for $\psi_0$, one multiplies $\psi_2$ to Eq.(\ref{NP3}) and makes use of the identities $D\psi_2=3\rho\psi_2,\; \delta\psi_2=3\tau\psi_2$ which are satisfied by background type D spacetime, giving rise to
\be
\tilde{E}_2(\psi_2\sigma)-\tilde{E}_1(\psi_2\kappa)=\psi_2\psi_0
\en
where
\begin{equation}
\begin{aligned}
\tilde{E}_1&=\delta-4\tau+\pi^*-\alpha^*-3\beta,\\
\tilde{E}_2&=D-4\rho-\rho^*-3\varepsilon+\varepsilon^*,
\end{aligned}
\end{equation}
Operating $\tilde{E}_1, \tilde{E}_2$ onto Eqs.(\ref{NP1}) and (\ref{NP2}) respectively, and then subtracting one equation from the other, one has
\be\label{NPf0}
(\tilde{E}_2F_2-\tilde{E}_1F_1-3\psi_2)\psi_0=T_0.
\en
Note that the commutation relation $\tilde{E}_2 J_2-\tilde{E}_1 J_1=0$ has been used here. The full set of NP equations are invariant under the interchange $l^\mu\leftrightarrow n^\mu, m^\mu\leftrightarrow \bar{m}^\mu$, applying this interchange to Eq.(\ref{NPf0}) one has
\be\label{NPf1}
(\tilde{E}_4F_4-\tilde{E}_3F_3-3\psi_2)\psi_4=T_4,
\en
with
\begin{equation}
\begin{aligned}
F_3&=\delta+4\beta-\tau,  \quad\quad\quad \quad\quad \quad\;\; F_4=D+4\varepsilon-\rho,\\
\tilde{E}_3&=\delta^*+3\alpha+\beta^*+4\pi-\tau^*, \quad  \tilde{E}_4=\Delta_0+4\mu+\mu^*+3\gamma-\gamma^*.
\end{aligned}
\end{equation}
As expected, now Eqs.(\ref{NPf0}) and (\ref{NPf1}) are the decoupled equations for the curvature tensors $\psi_0$ and $\psi_4$, respectively.

For the specific background spacetime (\ref{DMBH}), one calculates the spin coefficients and derivative operators and substitute them into the perturbation equation (\ref{NPf1}), yields
\begin{equation}
\begin{aligned}
&\left[\frac{(r^2+a^2)^2}{\Delta}-a^2\sin^2\theta\right]\frac{\partial^2 \psi}{\partial t^2}+\frac{4Mar-2af(r)+2 a r^2}{\Delta}\frac{\partial^2 \psi}{\partial t\partial\varphi}+\left[\frac{a^2}{\Delta}-\frac{1}{\sin^2\theta}\right]\frac{\partial^2 \psi}{\partial \varphi^2}\\
&-\Delta^2\frac{\partial}{\partial r}\left(\Delta^{-1}\frac{\partial \psi}{\partial r}\right)+4\left[\frac{M (r^2-a^2)+(r^2+a^2)(f'(r)/2-r)-r f(r)+r^3}{\Delta}-r-i a \cos\theta\right]\frac{\partial\psi}{\partial t}\\
&-\frac{1}{\sin\theta}\frac{\partial}{\partial\theta}
\left(\sin\theta\frac{\partial \psi}{\partial\theta}\right)+4\left[\frac{a(f'(r)/2-M)}{\Delta}+\frac{i\cos\theta}{\sin^2\theta}\right]\frac{\partial\psi}{\partial\varphi}+(4\cot^2\theta+2-f''(r)/2+1)\psi=4\pi\Sigma T,\label{Teukolskyeq}
\end{aligned}
\end{equation}
where $f(r)=r^2 \exp\left(-\frac{2\tilde{k} r^{\tilde{\sigma}-2} (r_c+r)^{\tilde{\sigma}-2}}{r_c(\tilde{\sigma}-2)}\right)$, and $\psi\equiv\psi_4/\rho^4$ is the gravitational perturbation with helicity $s=-2$.

The homogeneous equation can be separated by writing the gravitational perturbation $\psi$ as
\be\label{sepv}
\psi=e^{-i\omega t}e^{i m\varphi}S(\theta) R(r),
\en
substituting (\ref{sepv}) into Eq.(\ref{Teukolskyeq}), we obtain the radial and angular equations respectively as
\begin{align}
&\frac{(r^2+a^2)^2}{\Delta}\omega^2 R(r)-\frac{4 M a r-2af(r)+2 a r^2}{\Delta}\omega m  R(r)+\frac{a^2}{\Delta} m^2 R(r)\nonumber\\
&+\Delta^2\frac{\partial}{\partial r}\left(\Delta^{-1}\frac{\partial R(r)}{\partial r}\right)-4\left[\frac{a(f'(r)/2-M)}{\Delta}\right] i m R(r)\nonumber\\
&+4i\omega\left[\frac{M (r^2-a^2)+(r^2+a^2)(f'(r)/2-r)-r f(r)+r^3}{\Delta}-r\right]R(r)+(f''(r)/2-1-\lambda)R(r)=0,\label{radialeq}\\
&\frac{1}{\sin\theta}\frac{d}{d\theta}\left(\sin\theta\frac{dS(\theta)}{d\theta}\right)+\left[a^2\omega^2\cos^2\theta+4a\omega\cos\theta-
\left(\frac{m^2-4m\cos\theta+4}{\sin^2\theta}\right)+\mathcal{E}_{lm}\right]S(\theta)=0,\label{angularEq}
\end{align}
The solutions to the angular sector equation (\ref{angularEq}) are the spin-weighted spheroidal harmonics. When $a\omega=0$, the solution to Eq.(\ref{angularEq}) reduce to the spin-weighted spherical harmonics. In Eq.(\ref{radialeq}), $\lambda=\mathcal{E}_{lm}-2 a m \omega+a^2\omega^2-2$, here $\mathcal{E}_{lm}$ is the eigenvalue of the spherical harmonics.
The radial equation (\ref{radialeq}) can be rewritten in a simplified form through introducing a potential function
\be\label{radialeq2}
\Delta^2\frac{\partial}{\partial r}\left(\Delta^{-1}\frac{\partial R(r)}{\partial r}\right)-V(r)R(r)=0
\en

By examining the behaviors of potential $V(r)$ at $r\rightarrow r_{+}$ and $r\rightarrow \infty$, it's found the homogenous Teukolsky equation admit two branches of asymptotic solutions
\begin{equation}
\begin{aligned}
&R^H_{lm\omega}=B^{hole}_{lm\omega}\Delta^2e^{-i p_{m\omega} r^*},\quad\quad\quad r\rightarrow r_{+}\\
&R^H_{lm\omega}=B^{out}_{lm\omega}r^3e^{i \omega r^*}+\frac{B^{in}_{lm\omega}}{r}e^{-i \omega r^*},\quad\quad r\rightarrow \infty\\
\label{asymp11}
\end{aligned}
\end{equation}
and
\begin{equation}
\begin{aligned}
&R^\infty_{lm\omega}=D^{out}_{lm\omega}e^{i p_{m\omega} r^*}+D^{in}_{lm\omega}\Delta^2e^{-i p_{m\omega} r^*},\quad\quad\quad r\rightarrow r_{+}\\
&R^\infty_{lm\omega}=D^{\infty}_{lm\omega}r^3e^{i \omega r^*},\quad\quad r\rightarrow \infty\\
\label{asymp12}
\end{aligned}
\end{equation}
where $p_{m\omega}=\omega-m\omega_{+}$, $\omega_{+}=a/(2\mathcal{M}(r_{+})r_{+})$.  For the inhomogeneous Teukolsky equation with source term $\mathcal{T}(r)$, the general solution can be given using the theory of Green's function as
\be\label{TeuksolGreen}
R_{lm\omega}(r)=Z^H_{lm\omega}(r)R^\infty_{lm\omega}(r)+Z^\infty_{lm\omega}(r)R^H_{lm\omega}(r)
\en
with
\begin{equation}
\begin{aligned}\label{ZHZinf}
Z^H_{lm\omega}(r)&=\frac{1}{2i\omega B^{in}_{lm\omega}}\int_{r_{+}}^r dr'\frac{R^H_{lm\omega}(r')\mathcal{T}_{lm\omega}(r')}{\Delta(r')^2},\\
Z^\infty_{lm\omega}(r)&=\frac{B^{hole}_{lm\omega}}{2i\omega B^{in}_{lm\omega}D^\infty_{lm\omega}}\int_{r_{+}}^r dr'\frac{R^\infty_{lm\omega}(r')\mathcal{T}_{lm\omega}(r')}{\Delta(r')^2}.
\end{aligned}
\end{equation}
By construction one has $Z^H_{lm\omega}(r_{+})=Z^\infty_{lm\omega}(\infty)=0$, so the solution (\ref{TeuksolGreen}) is purely ingoing at the horizon and purely outgoing at infinity, it's physically reasonable. The source term $\mathcal{T}(r)$ describes the secondary in the EMRI system, which is viewed as a mass point, the specific form of $\mathcal{T}(r)$ can be found in \cite{Hughes:1999bq}.

\subsection{Sasaki-Nakamura Equation}
As shown above, the asymptotic solution of the radial equation diverges at infinity. By examining the asymptotic behavior of the potential function $V(r)$ at $r\rightarrow\infty$, it is noted that the long-rangeness of the potential $V(r)$ causes the divergence of the solution at infinity. This obstructs the numeric resolution of the Teukolsky equation. To facilitate solving the Teukolsky equation with numeric method, a coordinate transformation should be performed to obtain potential functions being of short-range characteristics, so that the asymptotic solutions are free from divergence at infinity, allowing to impose appropriate boundary conditions at infinity. To obtain the Teukolsky equation of short-rangeness, we perform the SN transformation\cite{Hughes:1999bq,Sasaki:1981sx}
\be
R=\frac{1}{\eta}\left[\left(\alpha+\frac{\beta'}{\Delta}\right)\chi-\frac{\beta}{\Delta}\chi'\right],\label{solR}
\en
In the above transformation, we choose $\alpha$, $\beta$ to be
\begin{equation}
\begin{aligned}\label{alphabeta}
\beta&=\Delta\left(-2 i K(r)+\frac{d\Delta(r)}{dr}-\frac{4\Delta(r)}{r}\right),\\
\alpha&=-i \frac{K(r)}{\Delta(r)}\beta+3 i \frac{d K(r)}{dr}+\lambda+\frac{6\Delta}{r^2}+1-\frac{1}{2}\frac{d^2\Delta(r)}{dr^2},
\end{aligned}
\end{equation}
where $K(r)=(r^2+a^2)\omega-a m$. Further introducing
\be\label{Xchi}
X=\frac{\sqrt{r^2+a^2}}{\Delta}\chi,
\en
then Eq.(\ref{radialeq}) can be cast to the form
\be
\frac{d^2X}{dr^{*2}}-\mathcal{F}\frac{d X}{dr^{*}}-\mathcal{U}X=0,\label{radialeqX}
\en
which is called the SN equation. Therein, the function $\mathcal{F}$ is given by
\be
\mathcal{F}=\frac{1}{\eta}\frac{d\eta(r)}{dr}\frac{\Delta(r)}{r^2+a^2},
\en
with
\be
\eta=c_0+\frac{c_1}{r}+\frac{c_2}{r^2}+\frac{c_3}{r^3}+\frac{c_4}{r^4},
\en
where the coefficients $c_i$ are given by
\begin{equation}
\begin{aligned}
c_0&=2\lambda + \lambda^2 + 12 a m \omega - 12 i M \omega - 12 a^2 \omega^2,\\
c_1&=-8 i \left( -a m  \lambda - 6 \tilde{k}\tilde{\sigma} \omega/r_c - 3 a^2  \omega + a^2  \lambda \omega \right),\\
c_2&=12 \left( 2 \tilde{k}\tilde{\sigma}/r_c + a^2 - 2 a^2 m^2  + 2 i a m M - 10 i \tilde{k}\tilde{\sigma}  \omega + 4 a^3 m  \omega - 2 i a^2 M  \omega - 2 a^4  \omega^2 \right),\\
c_3&=24 i \left( -2 a \tilde{k}\tilde{\sigma} m/r_c + 2 i \tilde{k}\tilde{\sigma} M/r_c+ 10 i \tilde{k}\tilde{\sigma} /3 - a^3 m + i a^2 M  \right.\\
&\left.\;\;\;+ 10 \tilde{k}^2\tilde{\sigma}^2\omega/r_c^2 + 2 a^2 \tilde{k}\tilde{\sigma} \omega/r_c + a^4  \omega + 10 \tilde{k}\tilde{\sigma} r_c \omega+5(1-\tilde{\sigma})\tilde{k}\omega/r_c \right),\\
c_4&= 228 \tilde{k}^2\tilde{\sigma}^2/r_c^2 + 48 a^2 \tilde{k}\tilde{\sigma}/r_c + 12 a^4 + 100 i a \tilde{k}\tilde{\sigma} m  + 180 \tilde{k} \tilde{\sigma}M  + 180 \tilde{k} \tilde{\sigma}r_c \\
&\;\;\;- 840 i \tilde{k}^2\tilde{\sigma}^2 \omega/r_c - 100 i a^2 \tilde{k} \tilde{\sigma}\omega - 420 i \tilde{k}\tilde{\sigma} r_c^2 \omega +180(1-\tilde{\sigma})\tilde{k}/(2r_c)-420i(1-\tilde{\sigma})\tilde{k}\omega.
\end{aligned}
\end{equation}
The potential function $\mathcal{U}$ in Eq.(\ref{radialeqX}) is given by
\be
\mathcal{U}=\frac{\Delta U_1}{(r^2+a^2)^2}+G^2+\frac{\Delta}{r^2+a^2}\frac{d G}{dr}-\mathcal{F}G,
\en
where
\begin{equation}
\begin{aligned}
G&=-\frac{1}{r^2+a^2}\frac{d\Delta}{dr}+\frac{r\Delta}{(r^2+a^2)^2},\\
U_1&=V(r)+\frac{\Delta^2}{\beta}\left[\frac{d}{dr}\left(2\alpha+\frac{d\beta/dr}{\Delta}\right)-\frac{d\eta/dr}{\eta}\left(
\alpha+\frac{d\beta/dr}{\Delta}\right)\right].
\end{aligned}
\end{equation}

The potential functions $\mathcal{F}$ and $\mathcal{U}$ are readily confirmed to be of short-rangeness, satisfying $\mathcal{F}=O(r^{*-n})$ and $\mathcal{U}=-\omega^2+O(r^{*-n})$ ($n\geq 2$) when $r^*\rightarrow \pm\infty$.
Thus, the SN equation (\ref{radialeqX}) admits the following asymptotic solutions:
\begin{equation}
\begin{aligned}
&X^H_{lm\omega}=e^{-i p_{m\omega} r^*}, \quad\quad\quad\quad\quad\quad r\rightarrow r_{+}\\
&X^H_{lm\omega}=A^{out}_{lm\omega}\bar{P}(r)e^{i \omega r^*}+A^{in}_{lm\omega}P(r)e^{-i \omega r^*},\quad\quad r\rightarrow \infty\\
\label{asymp21}
\end{aligned}
\end{equation}
and
\begin{equation}
\begin{aligned}
&X^\infty_{lm\omega}=C^{out}_{lm\omega}e^{i p_{m\omega} r^*}+C^{in}_{lm\omega}e^{-i p_{m\omega} r^*}, \quad\quad r\rightarrow r_{+}\\
&X^\infty_{lm\omega}=\bar{P}(r)e^{i \omega r^*},\quad\quad\quad\quad\quad\quad r\rightarrow \infty\\
\label{asymp22}
\end{aligned}
\end{equation}
where
\be
P(r)=1+\frac{A}{\omega r}+\frac{B}{(\omega r)^2}+\frac{C}{(\omega r)^3}
\en
with
\begin{equation}
\begin{aligned}
A&=-\frac{1}{2} i \left( 2+ \lambda + 2 a m \omega \right),\\
B&=-\frac{1}{8} \left( 2 \lambda + \lambda^2 + 4 a m \omega + 4 a m \lambda \omega + 4 a^2 m^2 \omega^2 + 4 i \left( -3 M \omega + 2 a m M \omega^2 \right) \right),\\
C&=\frac{1}{48} \left( -12 \left( 4 M  \omega - M  \lambda \omega - 2 a m M  \omega^2 + 2 a m M \lambda \omega^2 + 4 a^2 m^2 M  \omega^3 \right)\right.\\
 &\left.+ i \left( -8  \lambda - 2  \lambda^2 + \lambda^3 - 48 a m \omega - 8 a m \lambda \omega + 6 a m \lambda^2 \omega - 96 \tilde{k}\tilde{\sigma} \omega^2/r_c - 24 a^2  \omega^2 \right.\right.\\
 &\left.\left. + 8 a^2 \lambda \omega^2 + 12 a^2 m^2 \lambda \omega^2 + 32 a \tilde{k}\tilde{\sigma} m \omega^3/r_c + 16 a^3 m \omega^3 + 8 a^3 m^3 \omega^3 - 64 a m M^2 \omega^3 \right) \right).
\end{aligned}
\end{equation}
The coefficients $A$ and $B$ are identical with that of Kerr black hole, while the coefficient $C$ differ from that of Kerr black hole. When $\tilde{k}\rightarrow0$, $C$ here reduces to the Kerr case. It's evident that the asymptotic solutions of master equation are regular at infinity now. The short-range nature of the potential functions $\mathcal{F}$ and $\mathcal{U}$ allows to solve the SN equation numerically, we impose boundary conditions through the asymptotic solutions (\ref{asymp21}) and (\ref{asymp22}) and implement numeric integrations to obtain $X^{H,\infty}_{lm\omega}(r)$ at any location. Then, by applying the inverse transformations of (\ref{solR}) and (\ref{Xchi}), the solution to the radial Teukolsky equation (\ref{radialeq}) or (\ref{radialeq2}) can be obtained.

\section{Solve the Master Equations\label{section3}}

\subsection{Orbit Equation}
The background spacetime we considered in this paper is of Petrov type D, it possesses sufficient symmetries to allow for integrability of the motion of the secondary. In this case, it's convenient to construct the orbit equations of the secondary in EMRIs with the Hamilton-Jacobi formulation, which gives rise to
\begin{equation}
\begin{aligned}
\Sigma \frac{dr}{d\tau} &= \pm \sqrt{V_r},\\
\Sigma \frac{d\theta}{d\tau} &= \pm \sqrt{V_{\theta}},\\
\Sigma \frac{d\phi}{d\tau} &=  V_\phi, \\
\Sigma \frac{dt}{d\tau} &=  V_t,\label{geodesic}
\end{aligned}
\end{equation}
with
\begin{equation}
\begin{aligned}\label{geodpot}
V_r &= \left[ E(r^2 + a^2) - L_z a \right]^2 -
\Delta\left[\mu^2 r^2 + (L_z-aE)^2 + Q\right],\\
V_{\theta} &=  Q -\cos^2{\theta}\left[  a^2(\mu^2-E^2)
+ \frac{L_z^2}{\sin^2{\theta}} \right],\\
V_{\phi} &= \frac{L_z}{\sin^2{\theta}} - aE
+ \frac{a}{\Delta}\left[  E(r^2+a^2) -L_za \right],\\
V_t &= a \left(L_z - aE\sin^2{\theta}\right)
+ \frac{r^2+a^2}{\Delta} \left[ E(r^2+a^2) - L_za \right],
\end{aligned}
\end{equation}
where $\mu$, $E$ and $L_z$ are the mass, energy and angular momentum of the secondary, $Q$ is the Carter constant.

For bound orbits, there exist divergence at the turning points $\theta_{min}$ and  $\theta_{max}$, which obstructs the numerical implementation. To avoid the divergence, it's useful to introduce $\chi$ as $z=\cos^2\theta=z_{-}\cos^2\chi$\cite{Babak:2006uv}, where $z_{-}$ is given by
\be
\beta(z_{+}-z)(z_{-}-z)
=   \beta z^{2}-z\left[Q+L_{z}^{2}+a^{2}(\mu^2-E^{2})\right] +Q,
\en
with $\beta=a^{2}(\mu^2-E^{2})$. When $\theta$ goes from one turning point $\theta_{min}$ to the other $\theta_{max}$ and back to $\theta_{min}$, the parameter $\chi$ ranges from 0 to $2\pi$.  The equation of motion for $\theta$ above can then be rewritten with $\chi$ as
\be\label{eqchi}
\frac{d \chi}{d t} =\frac{\sqrt{\beta\,\left[z_{+} -
z(\chi)\right]}}{\gamma + a^{2}\,E\,z(\chi)},
\en
where $\gamma = E[\left(r^2+ a^2\right)^{2}/\Delta - a^2 ]-2 \mathcal{M}(r) r a L_{z}/\Delta$. Now the equation is free of divergence, numerical integration can be implemented successfully. Solving for $\chi$ and substituting it into the equation of $\phi$ in (\ref{geodesic}) and (\ref{geodpot}) enable us to integrate out $\phi$ numerically.

In this paper, we consider the inclined circular orbit. For the inclined circular orbit, the motion is periodic with well-defined frequencies. Since the motions of $\theta$ and $\phi$ are separately periodic, we need to determine the periods $T_\theta$, $T_\phi$ and the associated frequencies $\Omega_\theta$, $\Omega_\phi$ separately. These frequencies serve as the constituents of the harmonic frequencies defined below. From (\ref{eqchi}) we have
\be
t_0(\chi)=\int_0^\chi d\chi' \frac{\gamma + a^{2}\,E\,z(\chi')}{\sqrt{\beta\,\left[z_{+} -
z(\chi')\right]}},
\en
so the period of $\theta$ motion is
\be
T_\theta=4 t_0(\pi/2)= \frac{1}{\sqrt{(z_{+} - z_{-})\beta}}  \left( \mathrm{E}  \left( (z_{-} - z_{+}) \mathrm{E}\left(z_{-}/(z_{-} - z_{+})  \right) + z_{+} \mathrm{K}\left( z_{-}/(z_{-} - z_{+}) \right) \right) + 4 \mathrm{K}\left( z_{-}/(z_{-} - z_{+}) \right) \gamma \right),
\en
where $\mathrm{E}$ and $\mathrm{K}$ are the complete elliptic integral of the first and second kinds respectively. The frequency of the $\theta$ motion is given by $\Omega_\theta=2\pi/T_\theta$.

For the $\phi$ motion, we combine Eqs.(\ref{geodesic}), (\ref{geodpot}) and (\ref{eqchi}) to obtain
\be\label{eqphi}
\frac{d\phi}{d\chi}=\frac{1}{\sqrt{\beta(a_{+}-z)}}\left(\frac{L_z}{1-z}-\tilde{\delta}\right),
\en
where
\be
\tilde{\delta}=a E \left( \frac{r^2 + a^2}{\Delta} - 1 \right) - \frac{a^2 L_z}{\Delta}.
\en
Integrating Eq.(\ref{eqphi}), we have the azimuthal angle $\Phi$ the secondary go through in a $\theta$ period $T_\theta$.  The azimuthal angle is given by
\begin{equation}
\begin{aligned}
\Phi&=\frac{2}{(-1 + z_{-}) \sqrt{z_{+} \beta}}\left\{\sqrt{\frac{z_{+}}{z_{+}-z_{-}}} \left(-2 L_z + (-2 + z_{-}) \tilde{\delta}\right) \Pi\left(\frac{z_{-}}{-1 + z_{-}}, \frac{z_{-}}{z_{-} - z_{+}}\right)\right.\\
&\left.-\frac{2 i \sqrt{z_{+} z_{-}} \tilde{\delta}}{ \sqrt{z_{+}}-\sqrt{z_{-}} }\left[\sqrt{z_{-}} \cdot\mathrm{F}\left(i \cdot \text{arcsinh}\left(\sqrt{\frac{\sqrt{z_{+}} - \sqrt{z_{-}}}{\sqrt{z_{-}} + \sqrt{z_{+}}}}\right), \frac{(\sqrt{z_{-}} + \sqrt{z_{+}})^2}{(\sqrt{z_{-}} - \sqrt{z_{+}})^2}\right)\right.\right.\\
&\left.\left.+(z_{-}-2)\Pi\left(\frac{(-1 + \sqrt{z_{-}})(\sqrt{z_{-}} + \sqrt{z_{+}})}{(1 + \sqrt{z_{-}})(\sqrt{z_{-}} - \sqrt{z_{+}})}, i \cdot \text{arcsinh}\left(\sqrt{\frac{\sqrt{z_{+}} - \sqrt{z_{-}}}{\sqrt{z_{-}} + \sqrt{z_{+}}}}\right), \frac{(\sqrt{z_{-}} + \sqrt{z_{+}})^2}{(\sqrt{z_{-}} - \sqrt{z_{+}})^2}\right)\right.\right.\\
&\left.\left.
-(z_{-}-2)\Pi\left(\frac{(1 + \sqrt{z_{-}})(\sqrt{z_{-}} + \sqrt{z_{+}})}{(-1 +\sqrt{z_{-}})(\sqrt{z_{-}} - \sqrt{z_{+}})}, i \cdot \text{arcsinh}\left(\sqrt{\frac{\sqrt{z_{+}} - \sqrt{z_{-}}}{\sqrt{z_{-}} + \sqrt{z_{+}}}}\right), \frac{(\sqrt{z_{-}} + \sqrt{z_{+}})^2}{(\sqrt{z_{-}} - \sqrt{z_{+}})^2}\right)\right]\right\}.
\end{aligned}
\end{equation}
Hence, the frequency of $\phi$ motion is given by $\Omega_\phi=\Phi/T_\theta$. In terms of $\Omega_\theta$ and $\Omega_\phi$, a set of harmonic frequencies can be defined as
\be
\omega_{mk}=m\Omega_\phi+k\Omega_\theta,
\en
with which $Z^H_{lm\omega}$ and $Z^\infty_{lm\omega}$ are decomposed as
\begin{equation}
\begin{aligned}
Z^H_{lm\omega}&=\sum_k Z^H_{lmk}\delta(\omega-\omega_{mk}),\\
Z^\infty_{lm\omega}&=\sum_k Z^\infty_{lmk}\delta(\omega-\omega_{mk}).
\end{aligned}
\end{equation}
The coefficients $Z^H_{lmk}$ and $Z^\infty_{lmk}$ fully determine the energy and angular momentum fluxes.

\subsection{Backreaction of GWs}

Due to gravitational radiation, the EMRIs lose energy and angular momentum, causing the orbit of the secondary to evolve. Hence, working out the loss of energy and angular momentum enables to determine the orbit of the secondary with high-accuracy. The energy and angular momentum flow either to infinity or down to the horizon of the SBH. The energy flux at future null infinity can be calculated with Isaacson stress-energy tensor\cite{Thorne:1980ru}, which yields
\be
\left(\frac{dE}{d\Omega dt}\right)^{rad}_{r\rightarrow\infty}=\frac{1}{16\pi}\langle\dot{h}_{+}^2+\dot{h}_{-}^2\rangle,
\en
where the dot denotes derivative with respect to $t$, $\langle, \rangle$ denotes averaging over several wavelengths. By examining the large $r$ behavior of $\psi_4$, we obtain the energy and angular momentum fluxes at infinity, which can be expanded in terms of the harmonic components as
\begin{equation}
\begin{aligned}
\left(\frac{dE}{dt}\right)^{rad}_{r\rightarrow\infty}&=\sum_{lmk}\frac{|Z^H_{lmk}|^2}{4\pi\omega_{mk}^2},\\
\left(\frac{dL_z}{dt}\right)^{rad}_{r\rightarrow\infty}&=\sum_{lmk}\frac{m|Z^H_{lmk}|^2}{4\pi\omega_{mk}^3}.
\end{aligned}
\end{equation}

The energy flux and angular momentum flux at the horizon induced by GWs can be calculated using Hawking and Hartle's method\cite{Hawking:1972hy}.
Realizing that the energy carried by GWs radiated down to the black hole horizon causes an increase in the horizon area, hence Hawking and Hartle proposed the energy flux can be computed by examining the increase in the area of the event horizon
\begin{equation}
\label{energyloss1}
\frac{d^2E}{dtd\Omega}=\frac{1}{8\pi C_A}\frac{d^2A}{dtd\Omega},
\end{equation}
where for the DMBH we have
\be
C_A=\frac{\partial \mathcal{M}(r_{+})}{\partial M}+\left(\frac{\partial \mathcal{M}(r_{+})}{\partial r_{+}}+\mathcal{M}(r_{+})\right) \left(\frac{\partial r_{+}}{\partial
M}+\frac{m}{\omega_{mk}}\frac{\partial r_{+}}{\partial J} \right),
\en
where $J=a M$. The increase in the area of the black hole's event horizon can be calculated by integrating the congruence of null geodesics over the event horizon. The integration of the convergence of the null geodesics generators over the horizon gives rise to the rate of change of the horizon area. The gravitational perturbation causes the perturbations of the convergence and shear generators of the null geodesics at the horizon, these perturbations determine the rate of change of horizon area
\be\label{dA}
\frac{d^2A}{dtd\Omega}=\frac{2\mathcal{M}(r_{+})r_{+}}{\epsilon}|\sigma^{HH}|^2,
\en
here we have $\epsilon=(f'(r_{+})-2M)/(4(r_{+}^2+a^2)^2)$, $f(r)$ is given below Eq.(\ref{Teukolskyeq}). $\sigma^{HH}$ is the spin coefficient or shear generator $\sigma$ in Hawking-Hartle null tetrads, $\sigma^{HH}$ is related to the field component $\psi_0$ through the NP equation \cite{Hawking:1972hy}.
From the asymptotic form of the field components at the horizon
\begin{equation}
\begin{aligned}
\Upsilon_2&=\psi_0\sim\exp(-i\omega t+im\phi) \;_2S_{lm}(\theta)Y^\infty\Delta^{-2}e^{-ip_{mk}r^*},\\
\Upsilon_{-2}&=(r-ia\cos\theta)^4\psi_4\sim\exp(-i\omega t+im\phi) \;_{-2}S_{lm}(\theta)Z^\infty\Delta^{2}e^{-i p_{mk}r^*},
\end{aligned}
\end{equation}
and the  Teukolsky-Starobinsky relation
\be
\mathscr{D}\mathscr{D}\mathscr{D}\mathscr{D}R_{-2}=\frac{1}{4}R_2,
\en
where $\mathscr{D}=\partial_r-\frac{i K(r)}{\Delta}$ and $K(r)$ is given below Eq.(\ref{alphabeta}), we obtain
\begin{equation}
\begin{aligned}\label{YZ}
CY^\infty&=256 p_{mk} r_{+} \mathcal{M}(r_{+})Z^\infty\left[
(-i+p_{mk} r_{+}-4ip_{mk}^2r_{+}^2+4p_{mk}^3r_{+}^3)\mathcal{M}(r_{+})^3-ir_{+}^3(\mathcal{M}'(r_{+})-1)^3\right.\\
&\left.+r_{+}^2(-3i+p_{mk}r_{+})\mathcal{M}(r_{+})(\mathcal{M}'(r_{+})-1)^2
+r_{+}(-3i+2p_{mk}r_{+}-4ip_{mk}^2r_{+}^2)\mathcal{M}^2(r_{+})(\mathcal{M}'(r_{+})-1)
\right].
\end{aligned}
\end{equation}
From the above equations (\ref{dA})-(\ref{YZ}), we obtain the energy flux in unit solid angle
\begin{equation}
\label{energyloss}
\frac{d^2E}{dtd\Omega}=\frac{{}_2S_{lm}^2(\theta)}{8\pi C_A}\frac{1}{16\epsilon(2\mathcal{M}(r_{+})r_{+})^3(k^2+4\epsilon^2)}\frac{C_1^2}{|C|^2}|Z^\infty|^2,
\end{equation}
with
\begin{equation}
\begin{aligned}
C_1=&256 p_{mk} r_{+} \mathcal{M}(r_{+})\left[
r_{+}^2 \left( -3 i + p_{mk} r_{+} \right) \mathcal{M}(r_{+}) \left( \mathcal{M}'(r_{+})-1  \right)^2- i r_{+}^3 \left( \mathcal{M}'(r_{+})-1 \right)^3 \right.\\
& \left.+\left( -i + p_{mk} r_{+} - 4 i p_{mk}^2 r_{+}^2 + 4 p_{mk}^3 r_{+}^3 \right) \mathcal{M}(r_{+})^3
+ r_{+} \left( -3 i + 2 p_{mk} r_{+} - 4 i p_{mk}^2 r_{+}^2 \right) \mathcal{M}(r_{+})^2 \left( \mathcal{M}'(r_{+})-1 \right)\right],\\
|C|^2=&\left((\lambda + 2)^2 + 4a\omega_{mk} m - 4a^2\omega_{mk}^2\right)
\left(\lambda^2 + 36a\omega_{mk} m - 36a^2\omega_{mk}^2\right)\\
&+ (2\lambda + 3)(96a^2\omega_{mk}^2 - 48a\omega_{mk} m)
+ 144\omega_{mk}^2(M^2 - a^2).
\end{aligned}
\end{equation}
The energy and angular fluxes flow into the horizon can then be given by summing up all harmonic modes as
\begin{equation}
\begin{aligned}
\left(\frac{dE}{dt}\right)^{rad}_{r\rightarrow r_{+}}&=\sum_{lmk}\alpha_{lmk}\frac{|Z^\infty_{lmk}|^2}{4\pi\omega_{mk}^2},\\
\left(\frac{dL_z}{dt}\right)^{rad}_{r\rightarrow r_{+}}&=\sum_{lmk}\alpha_{lmk}\frac{m|Z^\infty_{lmk}|^2}{4\pi\omega_{mk}^3},
\end{aligned}
\end{equation}
where the coefficients $\alpha_{lmk}$ can be read off straightforwardly from Eq.(\ref{energyloss})
\be
\alpha_{lmk}=\frac{\omega_{mk}^2C_1^2}{32C_A\epsilon (2\mathcal{M}(r_{+})r_{+})^3(p_{mk}^2+4\epsilon^2)^2|C|^2}.
\en
With the energy flux and angular momentum flux at infinity and at the horizon, the total energy flux and total angular momentum flux can be obtained, which are the sum between the two contributions from both the horizon and the infinity.

The loss of energy and angular momentum due to gravitational radiation induces the evolution of the orbit of the secondary in EMRIs. In this paper, we consider the inclined circular orbit. The circular orbit satisfies the conditions $\dot{V_r}=0$ and $\dot{V_r}'=0$, from which we obtain
\begin{equation}
\begin{aligned}\label{evolutionEq}
\dot{Q}&=\frac{c_{11}}{d}\dot{E}+\frac{c_{12}}{d}\dot{L}_z,\\
\dot{r}&=\frac{c_{21}}{d}\dot{E}+\frac{c_{22}}{d}\dot{L}_z,
\end{aligned}
\end{equation}
where
\begin{equation}
\begin{aligned}
c_{11}=&4 \left[ 4 E r \left( -a L_z + E (a^2 + r^2) \right)
- 2 r \mu^2 \left( a^2 + r^2 - 2 r \mathcal{M}(r) \right)
+ 2 \left( (-a E + L_z)^2 + Q + r^2 \mu^2 \right) \left( \mathcal{M}(r) + r \left( -1 + \mathcal{M}'(r) \right) \right) \right] \\
&\cdot \left[ E r (a^2 + 2 r^2) + a (a E - L_z) \mathcal{M}(r) + a (a E - L_z) r \mathcal{M}'(r) \right]  - 4 r \left[ E r (a^2 + r^2) + 2 a (a E - L_z) \mathcal{M}(r) \right] \\
& \cdot \left[ a^2 E^2 - L_z^2 - Q + 6 E^2 r^2 - a^2 \mu^2 - 6 r^2 \mu^2
+ 6 r \mu^2 \mathcal{M}(r)\right.\\
&\left.+ 2 \left( a^2 E^2 - 2 a E L_z + L_z^2 + Q + 3 r^2 \mu^2 \right) \mathcal{M}'(r)
+ r \mathcal{M}''(r) \left( (a E - L_z)^2 + Q + r^2 \mu^2 \right) \right],\\
c_{12}=& -4 \left[ 4 E r \left( -a L_z + E (a^2 + r^2) \right)
- 2 r \mu^2 \left( a^2 + r^2 - 2 r \mathcal{M}(r) \right)
+ 2 \left( (-a E + L_z)^2 + Q + r^2 \mu^2 \right)
\left( \mathcal{M}(r) + r \left( -1 + \mathcal{M}'(r) \right) \right) \right] \\
& \cdot \left[ (a E - L_z) \mathcal{M}(r) + r \left( L_z + (a E - L_z) \mathcal{M}'(r) \right) \right]  + 4 r \left[ L_z r + 2 (a E - L_z) \mathcal{M}(r) \right] \\
&\cdot \left[ a^2 E^2 - L_z^2 - Q + 6 E^2 r^2 - a^2 \mu^2 - 6 r^2 \mu^2
+ 6 r \mu^2 \mathcal{M}(r)\right.\\
&\left.+ 2 \left( a^2 E^2 - 2 a E L_z + L_z^2 + Q + 3 r^2 \mu^2 \right) \mathcal{M}'(r)
+ r \mathcal{M}''(r) \left( (a E - L_z)^2 + Q + r^2 \mu^2 \right) \right],\\
c_{21}=&4 \left( a^4 E - a^3 L_z - 2 a^2 E r^2 + a L_z r^2 - 3 E r^4 \right) \mathcal{M}(r)+4 r \left( a^2 + r^2 \right) \left[ E (a^2 + r^2) + \left( a^2 E - a L_z + E r^2 \right) \mathcal{M}'(r) \right],\\
c_{22}=&
-4a \left[ a L_z r + (a^2 E - a L_z - E r^2) \mathcal{M}(r) + r (a^2 E - a L_z + E r^2) \mathcal{M}'(r) \right],\\
d=&2 \left[ 4 E r \left( -a L_z + E (a^2 + r^2) \right)
- 2 r \mu^2 \left( a^2 + r^2 - 2 r \mathcal{M}(r) \right)
+ 2 \left( (-a E + L_z)^2 + Q + r^2 \mu^2 \right)
\left( \mathcal{M}(r) + r \left( -1 + \mathcal{M}'(r) \right) \right) \right] \\
& \cdot \left[ \mathcal{M}(r) + r \left( -1 + \mathcal{M}'(r) \right) \right] + 2 \left( a^2 + r^2 - 2 r \mathcal{M}(r) \right)  \left[ a^2 E^2 - L_z^2 - Q + 6 E^2 r^2 - a^2 \mu^2 - 6 r^2 \mu^2
+ 6 r \mu^2 \mathcal{M}(r)\right.\\
&\left.+ 2 \left( a^2 E^2 - 2 a E L_z + L_z^2 + Q + 3 r^2 \mu^2 \right) \mathcal{M}'(r)
+ r \mathcal{M}''(r) \left( (a E - L_z)^2 + Q + r^2 \mu^2 \right) \right].
\end{aligned}
\end{equation}
The evolution equations (\ref{evolutionEq}) together with the orbit equations (\ref{geodesic}) completely determine the orbital evolution of the secondary in EMRIs.

\subsection{Waveforms of GWs}

The source term describing a point particle on a circular orbit is given by
\be\label{source}
\mathcal{T}_{lm\omega}(r)=\int dt e^{i\omega t-im\phi}\{\left[A_{nn0}+A_{n\bar{m}0}+A_{\bar{m}\bar{m}0}\right]\delta(r-r_0)
+\partial_r\left([A_{n\bar{m}1}+A_{\bar{m}\bar{m}1}]\delta(r-r_0)\right)+\partial_r^2[A_{\bar{m}\bar{m}2}\delta(r-r_0)]\}
\en
where the definition of the quantities $A_{abi}$ can be found in \cite{Hughes:1999bq}. Substituting the source term into Eq.(\ref{ZHZinf}) and change the integration variable from $t$ to $\chi$, one has
\begin{equation}
\begin{aligned}\label{ZHZinfsol}
Z^H_{lmk}&=\frac{\pi}{i\omega_{mk}T_\theta B^{in}_{lmk}}\sum_{\pm}\int_0^\pi d\chi\frac{\gamma+a^2 E z(\chi)}{\sqrt{\beta[z_{+}-z(\chi)]}}
e^{\pm i [\omega_{mk}t(\chi)-m\phi(\chi)]}{}_{\pm} I_{lmk}^H[r_0,z(\chi)],\\
Z^\infty_{lmk}&=\frac{\pi c_0}{4i\omega_{mk}^3d_{lm\omega}T_\theta B^{in}_{lmk}}\sum_{\pm}\int_0^\pi d\chi\frac{\gamma+a^2 E z(\chi)}{\sqrt{\beta[z_{+}-z(\chi)]}}
e^{\pm i [\omega_{mk}t(\chi)-m\phi(\chi)]}{}_{\pm} I_{lmk}^\infty[r_0,z(\chi)],
\end{aligned}
\end{equation}
where
\begin{equation}
_{\pm}I^{H,\infty}_{lmk}[r_0,z(\chi)] =
R^{H,\infty}_{lm\omega}(r_0)
\left[A^{\pm}_{nn0} + A^{\pm}_{n\bar m0} + A^{\pm}_{\bar m\bar m0}\right]
- {dR^{H,\infty}_{lm\omega}\over dr}\biggr|_{r_0}
\left[A^{\pm}_{n\bar m1} + A^{\pm}_{\bar m\bar m1}\right]
+ {d^2R^{H,\infty}_{lm\omega}\over dr^2}\biggr|_{r_0}\,
A^{\pm}_{\bar m\bar m2}\;,
\label{eq:I_def}
\end{equation}
the definition of the quantities $A^{\pm}_{abi}$ can be found in Ref.\cite{Hughes:1999bq} as well.

\begin{figure}[h]
\begin{center}
\includegraphics[width=0.5\textwidth]{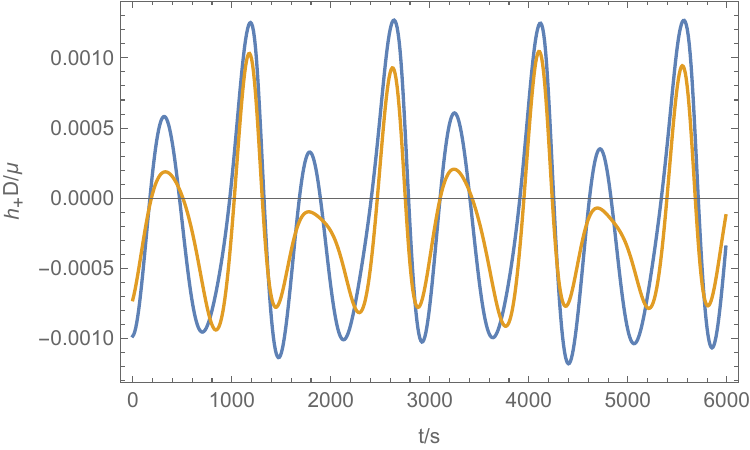}
\end{center}
\caption{Gravitational waveforms of EMRIs in Kerr (blue) and DMBH (orange).}\label{comp}.
\end{figure}

Now we have all the constituents to calculate the GWs emanating from EMRIs, for the calculations of this part we  follow the numerical algorithm given in Ref.\cite{Hughes:1999bq}.
The convergence of the asymptotic solutions of the SN equation given in Eqs.(\ref{asymp21}, \ref{asymp22}) at infinity allows us to determine the infinity using numeric method. Following the method given in \cite{Hughes:1999bq}, we implement a variant of Richardson extrapolation to compute the value of $X^\infty_{lm\omega}$ at infinity. We iterate until the successive difference of the sequence $X^\infty_{i;lm\omega}$ is less than $10^{-7}$, then the infinity used in numeric computations can be determined. This method is more fast and more accuracy than just using a very large number to denote infinity. The horizon radius used in numeric method can be determined with the similar extrapolation technique as well. After determining the horizon radius and infinity used in numeric method, one is able to impose the boundary conditions through the asymptotic solutions (\ref{asymp21}, \ref{asymp22}). Then the SN equation can be solved numerically and two branches solutions $X^H_{lmk}(r)$ and $X^\infty_{lmk}(r)$ are obtained, here $r$ denotes any location from $r_{+}$ to $\infty$. With the value of $X^H_{lmk}$ at infinity one can determine the undetermined coefficients $A^{in}_{lmk}$ and $A^{out}_{lmk}$ in Eq.(\ref{asymp21}).
Substituting $X^H_{lmk}$ and $X^\infty_{lmk}$ into Eq.(\ref{solR}) yields $R^H_{lmk}$ and $R^\infty_{lmk}$. Substituting all the results into Eq.(\ref{ZHZinfsol}) and performing numerical integration one obtains $Z^H_{lmk}$ and $Z^\infty_{lmk}$. With $Z^H_{lmk}$ and $Z^\infty_{lmk}$ one can compute the energy and angular momentum losses of the secondary and the evolution of the orbit. The gravitational waveform can be extracted through
\be
h_+(\theta,\phi,t) - i h_\times(\theta,\phi,t) =
\sum_{lmk} {1\over\omega_{mk}^2} Z^H_{lmk}
{_{-2}}S^{a\omega_{mk}}_{lm}(\theta) e^{i(m\phi - \omega t)}\;.
\label{waveform}
\en
We compute the spheroidal harmonics ${_{-2}}S^{a\omega_{mk}}_{lm}(\theta)$ and its eigenvalue $\mathcal{E}_{lm}$ with the package ``Spin-Weighted Spheroidal Harmonics'' in ``Black Hole Perturbation Toolkit''\cite{BHtoolkit}.  Fig.\ref{comp} shows the gravitational waveforms of EMRIs in Kerr and DMBH, the harmonic modes are truncated at $l=8$. The mass of the SBH is set to be $3\times10^6M_{\odot}$, and the mass of the secondary is set to be $5M_{\odot}$. The spin of the SBH $a$ is set to be 0.5, the inclined angle of the circular orbit $\iota$ is set to be $\frac{\pi}{6}$. As shown in Fig.\ref{comp}, the DM causes amplitude changes and phase shifts between the two waveforms, which allows to compare the two waveforms further by adjusting the DM parameters.

\section{Analysis of the Waveforms\label{section4}}

\subsection{Detector Response}
The strain of space-based GW detectors caused by GWs is given by
\begin{align}
    h=\frac{\sqrt{3}}{2}\left[F^{+}h_{+}+F^{\times}h_{\times}\right],\label{stainform}
\end{align}
where
$F^{+}$ and $F^{\times}$ in (\ref{stainform}) are the response functions. For LISA, the response functions are given by\cite{Barack:2003fp,Apostolatos:1994mx,Cutler:1997ta,Canizares:2012is}
\begin{equation}
\begin{aligned}
F^{+}=\frac{1}{2}(1+\cos^2\Theta)\cos(2\Phi)\cos(2\Psi)\\
-\cos(\Theta)\sin(2\Phi)\sin(2\Psi),\\
F^{\times}=\frac{1}{2}(1+\cos^2\Theta)\cos(2\Phi)\sin(2\Psi)
\\+\cos(\Theta)\sin(2\Phi)\cos(2\Psi).
\end{aligned}
\end{equation}
The direction of polarization $+$ is along $\hat{N}\times\hat{L}$, where $\hat{N}$ is the unit vector pointing from the detector to the GW source, $\hat{L}$ denotes the normal direction of the plane of the secondary's orbit. The axis of polarization state $\times$  is the one that is rotated counterclockwise from the $+$ axis by $45^{\circ}$ in the plane of the sky. The angles $(\Theta, \Phi)$ denote the latitude and azimuth in the detector-based coordinate system. The polarization angle $\Psi$ is the angle from the principal $+$ direction, $\pm\hat{N} \times
\hat{L}$, clockwise in the plane of the sky to the direction of
constant azimuth, $\pm \hat{N}\times (\hat{N}\times \hat{z})$, where $\hat{z}$ is the unit vector along the $z$ axis of the detector-based coordinate system. Since the satellites of LISA revolve around the sun, and the secondary revolves around the central SBH, the angles $\Theta, \Phi, \Psi$ are functions of time. By working out the relations between the detector-based coordinate system and the ecliptic-based system, one has
\begin{equation}
\begin{aligned}
\cos\Theta(t)&=\frac{1}{2}\cos\theta_S-\frac{\sqrt{3}}{2}\sin\theta_S\cos[\bar{\phi}_0+2\pi t/T-\phi_S]\\
\Phi(t)&=\bar{\alpha}_0+2\pi t/T\\
&+\tan^{-1}\left[\frac{\sqrt{3}\cos\theta_S+\sin\theta_S\cos[\bar{\phi}_0+2\pi t/T-\phi_S]}{2\sin\theta_S\sin[\bar{\phi}_0+2\pi t/T-\phi_S]}\right],
\end{aligned}
\end{equation}
where the angles ($\theta_S, \phi_S$) are latitude and azimuth of the source in the ecliptic-based system. $T=1$ year, and the constant angles $\bar{\phi}_0, \bar{\alpha}_0$ specify the initial orbital and rotational phase of the detector. The polarization angle $\Psi$ is given by

\begin{equation} \label{Psi}
\begin{aligned}
\tan\Psi&=\left\{\frac{1}{2}\cos\theta_L-\frac{\sqrt{3}}{2}
\sin\theta_L \cos[\bar\phi_0+2\pi(t/T)-\phi_L]\right.\\
&\left.-\cos\Theta(t)\left[
\cos\theta_L\cos\theta_S+\sin\theta_L\sin\theta_S\cos(\phi_L-\phi_S)
\right]\right\}/  \\
&\left[
\frac{1}{2}\sin\theta_L\sin\theta_S\sin(\phi_L-\phi_S)-\frac{\sqrt{3}}{2}\cos(\bar\phi_0+2\pi t/T)\cdot \right.\\
&\left.\left(\cos\theta_L\sin\theta_S\sin\phi_S
-\cos\theta_S\sin\theta_L\sin\phi_L\right)-\right.\\
&\left.\frac{\sqrt{3}}{2}\sin(\bar\phi_0+2\pi t/T)\left(\cos\theta_S\sin\theta_L\cos\phi_L\right.\right.\\
&\left.\left.-\cos\theta_L\sin\theta_S\cos\phi_S\right)\right].
\end{aligned}
\end{equation}
The angles $\theta_L, \phi_L$ denote the direction of $\hat{L}$ in the ecliptic-based system.

\begin{figure}[h]
\begin{center}
\includegraphics[width=0.5\textwidth]{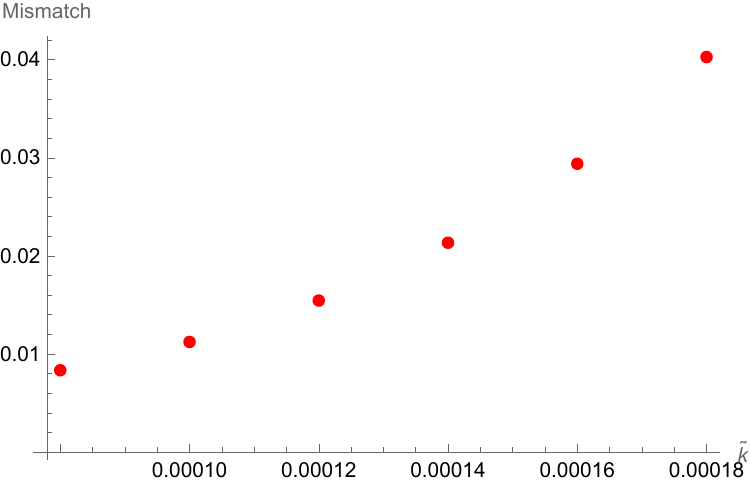}
\end{center}
\caption{Mismatch between the GWs of EMRIs in Kerr and DMBH for different mass parameter of DM halo $\tilde{k}$. The spin of the SBH $a$ is fixed to be $0.5$. }\label{misk1}.
\end{figure}

\subsection{Mismatch of Waveforms}
In this subsection, we compare the waveforms of Kerr and DMBH by computing the mismatch between them. The mismatch is defined with the inner product of two signals. A smaller mismatch value indicates that the waveforms of the two signals are more similar, while a larger mismatch value indicates a greater difference between them. If the mismatch is significantly larger than the detector noise threshold, it indicates the two waveforms are statistically distinguishable.
The inner product of two signals is defined as
\begin{eqnarray}
\langle x,h\rangle =  2 \int_0^{\infty} \frac{\tilde{x}(f)\tilde{h}^*(f) + \tilde{x}^*(f)
\tilde{h}(f)}{S_h(f)} df,\label{innerproduct}
\end{eqnarray}
where $\tilde{x}(f)$ and $\tilde{h}(f)$ are the frequency-domain representations of the time-domain signals $x(t)$ and $h(t)$
 \begin{eqnarray}
\tilde{x}(f) = \int^\infty_{-\infty} x(t)e^{-i2\pi tf}dt,\quad\quad\quad\quad\tilde{h}(f) = \int^\infty_{-\infty} h(t)e^{-i2\pi tf}dt.
\end{eqnarray}
The superscript ``$*$'' in Eq.(\ref{innerproduct}) denotes the complex conjugate of a quantity, and $S_h(f)$ in Eq.(\ref{innerproduct}) is the one-sided noise power spectral density of LISA \cite{Robson:2018ifk}, which is significant for surveying
the types of sources that can be detected by the LISA mission.

\begin{figure}[h]
\begin{center}
\includegraphics[width=0.5\textwidth]{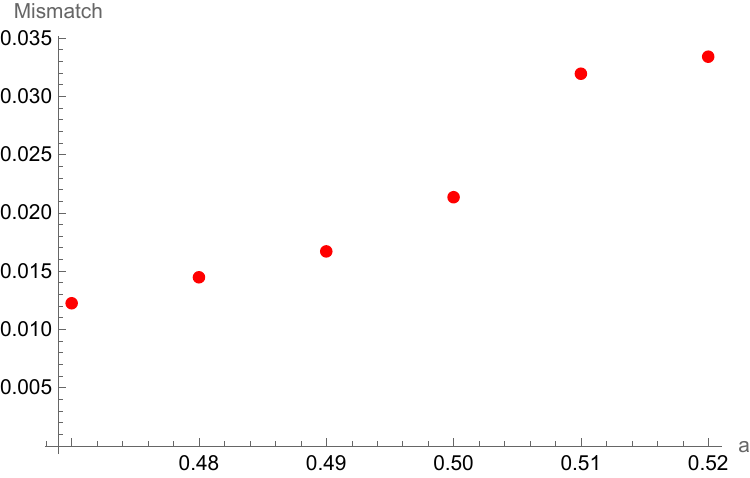}
\end{center}
\caption{Mismatch between the GWs of EMRIs in Kerr and DMBH for different spin of the SBH. The mass of the DM halo $\tilde{k}$ is fixed to be $1.4\times10^{-4}$.}\label{misa}.
\end{figure}

With the inner product defined in (\ref{innerproduct}), the overlap of two signals is given by
\be
\mathcal{O}=\frac{\langle x,h\rangle}{\sqrt{\langle x,x\rangle}\sqrt{\langle h,h\rangle}}.\label{overlap}
\en
The mismatch of two GW signals is defined as
\be
\mathscr{M}=1-\mathcal{O},
\en
with which the indistinguishable criterion of two
signals is given by \cite{Tan:2024utr,Mitman:2025tmj,
Zhao:2025sck,Drummond:2023wqc}
\begin{equation}
\mathscr{M}\leq \frac{N}{2\rho^2}.
\end{equation}
Where $\rho$ is the signal-to-noise ratio of the detector. For LISA, $\rho$ is taken to be 20\cite{Babak:2017tow}. $N$ is the number of the GW parameters, for DMBH $N=16$.  According to the indistinguishable criterion, two GW signals are distinguishable only when $\mathscr{M}>0.02$.
As shown in Fig.\ref{comp}, the DM halo causes amplitude changes and phase shifts between the gravitatinal waveforms of the EMRIs in DM black holes and Kerr black holes.
We adjust the mass parameter of the DM halo and compare the two types of gravitational waveforms. The result shows that, the amplitude of GWs decreases when the mass of the DM halo increases, and that the phase shift between the GWs of Kerr and DMBH increase when the mass of DM halo increases. Fig.\ref{misk1} shows that, for other parameters fixed, the mismatch increases with the mass parameter $\tilde{k}$ of the DM halo. As $\tilde{k}$ increases, the indistinguishable bound will be exceeded, e.g., if the spin parameter is fixed as $a=0.5$, when $\tilde{k}>1.3\times10^{-4}$ the indistinguishable criterion is exceeded, indicating the waveforms of Kerr and DMBH become distinguishable. Interestingly, the mismatch between the waveforms of the DMBH and the Kerr black hole is insensitive to whether the profile is core-like or cusp-like. For instance, when $a = 0.5$, for both the cored and the cuspy profiles of DM halo, the two waveforms become distinguishable for about $\tilde{k}>1.3\times10^{-4}$.
We adjusted the spin parameter of the SBHs. The result shows that for both Kerr and DMBH, the gravitational wave amplitude decreases when spin increases, and the phase shift between the two types of GWs increases with spin. As shown in Fig.\ref{misa},
the mismatch increases with spin.
If we expand the range of spin, calculations show that the mismatch does not vary monotonically with spin, but overall the mismatch still increases with the spin.
So the result indicates the larger the spin of SBH, the easier it is to distinguish the gravitational waveforms of Kerr and DMBH.

\section{Discussion and Conclusion\label{section5}}

In this paper, we compute the GWs emanating from EMRI systems in Kerr and rotating DM black holes using the Teukolsky method. We consider the profile of the DM halo to be of Dehnen type. We establish the perturbation equations for the curvature tensors $\psi_0$ and $\psi_4$ via the Teukolsky formalism. Substituting the DMBH metric into the perturbation equations and separating variables we obtain the radial and angular equations. The equation of angular sector is identical to that of the Kerr black hole, its solution is the spin-weighted spheroidal harmonics. The general solution to the radial inhomogeneous equation, which is physically reasonable, can be constructed formally with the solutions of the homogeneous equation using the theory of Green's function. We solve the radial equation with numerical method.  Due to the long-range nature of the potential functions in the radial equation, the asymptotic solutions diverge at infinity, which obstructs the numerical implementation. To solve the radial equation with numerical method, we perform the SN transformation to obtain the SN-type equation, the potential functions of which are short-ranged thus allow to be solved numerically. Considering the orbit of the secondary of EMRIs to be nearly circular, we compute the frequencies of the $\theta$-motion and $\phi$-motion, and thereby obtain the harmonic frequencies of the GWs. We also study the  back-reaction of gravitational radiation. We calculate the energy and angular momentum fluxes at infinity with the Isaacson stress-energy tensor.  We derive the fluxes into the horizon with Hawking's method by examining the increase in horizon area induced by GWs. The orbital evolution is then obtained by the total energy and angular momentum fluxes. We solve the SN equation numerically with Hughes' algorithm, with the solution to the SN equation we give the solution to the Teukolsky equation. Finally, the two GW polarizations $h_{+}$ and $h_{\times}$ are extracted by summing over the harmonic modes.

We compare the EMRI gravitational waveforms of Kerr and DMBH, and find changes in the amplitude and phase of the GWs induced by the DM halo. We consider the detector response, and calculate the strain of LISA produced by the GWs using the  response function of LISA. By computing the mismatch between the two types of waveforms, we find that the mismatch increases with the mass of the DM halo. When the mass parameter becomes sufficiently large, the indistinguishable criterion is exceeded, and the two waveforms become distinguishable. For instance, fixing $a=0.5$, when $\tilde{k}>1.3\times 10^{-4}$, the two waveforms can be distinguished. We adjust the black hole spin and calculate the GWs of EMRIs. The results show that the mismatch between the two waveforms increases with the spin of the SBH, indicating that the larger the spin, the easier it is to distinguish the gravitational waveforms of EMRIs in Kerr and in DMBH, and the more readily DM halos can be detected through GWs.

\section*{Acknowledgment}
This work is supported by Natural Science Foundation of Shandong Province Nos.ZR2023MA014.

\providecommand{\href}[2]{#2}\begingroup
\footnotesize\itemsep=0pt
\providecommand{\eprint}[2][]{\href{http://arxiv.org/abs/#2}{arXiv:#2}}

\end{document}